\documentclass[conference]{IEEEtran}
\IEEEoverridecommandlockouts
\usepackage{cite}
\usepackage{amsmath,amssymb,amsfonts}
\usepackage{algorithmic}
\usepackage{graphicx}
\usepackage{textcomp}
\usepackage{xcolor}
\usepackage{subcaption}
\def\BibTeX{{\rm B\kern-.05em{\sc i\kern-.025em b}\kern-.08em
    T\kern-.1667em\lower.7ex\hbox{E}\kern-.125emX}}
\begin{document}


\newcommand{\hatsuwaryo}{amount of speech production}
\newcommand{\hatsuwazikan}{speech duration}
\newcommand{\hukubudousa}{abdominal motion measurement device}
\newcommand{\hukubudousakeisoku}{abdominal motion measurement device}
\newcommand{\hukubudousakeisokudevice}{abdominal motion measurement device}
\newcommand{\suiteihatuwazikan}{estimated speech duration}
\newcommand{\realhatuwazikan}{actual time of speech}
\newcommand{\kuukiatsu}{air pressure}
\newcommand{\kasokudo}{acceleration}
\newcommand{\sanzikukasokudo}{3-axis acceleration}
\newcommand{\hatsuwaari}{Speech}
\newcommand{\hatsuwanashi}{No speech}
\newcommand{\hatsuwasikibetumodel}{speech discrimination model}
\newcommand{\bubunzikeiretudata}{partial time series data}
\newcommand{\kuukiatusensor}{barometric sensor}

\title{
Estimating Speech Duration by Measuring the Abdominal Movement Using a Barometric Sensor\\
}


\author{\IEEEauthorblockN{Rintaro Katagiri}
\IEEEauthorblockA{\textit{Depertment of Information Science and Technology} \\
\textit{Kyushu University}\\
Fukuoka, Japan \\
katagiri.rintaro.066@s.kyushu-u.ac.jp}
\and
\IEEEauthorblockN{Yutaka Arakawa}
\IEEEauthorblockA{\textit{Faculty of Information Science and Electrical Engineering} \\
\textit{Kyushu University}\\
Fukuoka, Japan \\
arakawa@ait.kyushu-u.ac.jp}
\and
\IEEEauthorblockN{Yugo Nakamura}
\IEEEauthorblockA{\textit{Faculty of Information Science and Electrical Engineering} \\
\textit{Kyushu University}\\
Fukuoka, Japan \\
y-nakamura@ait.kyushu-u.ac.jp}
}


\maketitle

\begin{abstract}


Measuring the amount of speech production in daily life is important for understanding communication in organizations and identifying mental disorders. However, measuring the amount of speech production can be problematic in terms of privacy.
We observed the whole-body condition during speech and noted that the abdomen strains during speech production. Therefore, we developed a less uncomfortable, inflatable abdominal motion measurement device using a barometric sensor to measure speech production indirectly. We measured speech production in 10 subjects and created a speech discrimination model using machine learning. However, the estimated speech duration in an actual meeting using this model was much longer than the actual duration. We found that the wearer's posture significantly affects the accuracy of the speech discrimination model developed in this study. We plan to improve the abdominal motion measurement device to minimize the effect of posture and achieve more accurate speech production measurement.

\end{abstract}


\section{Introduction}

In recent years, numerous studies have explored the monitoring of daily activities for health management and enhancement through the utilization of wearable devices promoting a healthy lifestyle. Our team has engineered several sensing devices, including the Waiston Belt\cite{waistonbelt}, a belt-type wearable device, a sensing chair named Census\cite{census}, and smart gloves\cite{smartglove}, all aimed at recording various daily activities. Building upon these developments, the present study concentrates on the analysis of conversation, a fundamental aspect of our daily interactions.


Effective communication is pivotal in human society, with conversation serving as a metric for assessing the extent of interpersonal communication. Studies indicate a correlation between amount of speech production and health; for instance, students exhibiting higher speech production demonstrate fewer depressive symptoms\cite{studentlife_2014} and are linked to stress\cite{studentlife_2019}.

Previous study employed a mobile application, StudentLife, to gauge subjects' daily activities, using the smartphone's microphone to measure speech production. However, this method raises privacy concerns, and accurate measurement is hindered by factors such as the presence of ambient noise and the voice of the conversation partner. Additionally, high battery consumption renders it unsuitable for prolonged measurements. Therefore, alternative methods, not reliant on microphones, are imperative for accurately assessing the volume of daily conversation over extended periods in real-life situations. A cost-effective wearable device capable of continuous daily measurement is ideal.


Conversely, the field of mechano-acoustic involves measuring vibrations within the body externally, with proposed devices attached to the throat\cite{mechano_2016} and chest\cite{mechano_2020}. These devices not only detect voice but also capture data on blood flow, heartbeat, gait, and other vibrations. The challenge lies in processing this data to isolate specific vibrations. While privacy concerns persist, the conceptual nature of these devices and their size make long-term measurements impractical.


Given this context, our study focuses on observing the entire body during speech, emphasizing the force exerted on the abdomen. Recognizing that identifying an abdominal motion measurement device during speech could enable the exclusive measurement of speech production without privacy concerns or microphone usage, we embarked on this study. While previous studies have not attempted to recognize speech from such a device, abdominal motion caused by normal breathing in the absence of speech has been studied, with methods for sensing respiration from abdominal motion measurement devices explored. Various techniques, including a capacitance pressure sensor\cite{respiration_belt} and an ultrasonic transducer\cite{kaji2018wearable}, have been proposed.





Consequently, we developed an inflatable abdominal motion measurement device utilizing a pressure-sensing method\cite{mizukusa2023}. This device incorporates a barometric sensor enclosed within a sealed, inflatable bag, functioning as a respiration sensing device characterized by its simple structure, low cost, and minimal discomfort during wear.


Air pressure sensor-based Human Activity Recognition (HAR) has been conducted and has been utilized for various applications, including detecting walking\cite{lin2016smart} and identifying falls\cite{barometricfall}.

Upon reviewing existing literature, we discovered previous studies on respiratory measurement using \kuukiatusensor. However, these methods often involved attaching the sensor to a face mask\cite{zhou2020accurate} or placing it on a desk for measurement\cite{chen2020pervasive}, with none specifically focusing on \hukubudousa.


In this study, we evaluated the effectiveness of the proposed system in measuring speech duration. Participants wore the abdominal motion measurement device we developed, and their abdominal motion data were recorded under two conditions: ``No speech'' and ``Speech'', while standing. Additionally, 3-axis acceleration data were concurrently measured. A speech discrimination model was constructed and validated using both sets of data. The obtained results revealed that the model based on air pressure data achieved an accuracy of 89.8\%, the model relying on 3-axis acceleration data achieved an accuracy of 90.5\%, and the model combining both data achieved an impressive accuracy of 94.6\%. This suggests that our proposed device is effective in accurately measuring speech duration.

\section{Related Work}
In this section, we discuss related study on speech and health and related techniques for measuring the amount of speech production.

\subsection{Related study on speech and health}

Wang et al.\cite{studentlife_2014} utilized StudentLife to gauge various aspects of subjects' activities, including walking, cycling, conversation, and sleep. Additionally, mental health questionnaires were collected to investigate the factors influencing college students' grades and class attendance. The study aimed to establish a connection between subjects' academic performance and the collected data. Notably, the study revealed a positive correlation between students' grades and the frequency of their conversations. The amount of conversation also exhibited variations over the semester, with an increase observed before exams.

Dasilva et al.\cite{studentlife_2019} conducted a stress survey on college students employing StudentLife. The survey measured four key elements: conversation, sleep, cell phone use, and location. The study delved into the timing of conversations throughout the day and found that stressed college students engaged in more conversations from 9 a.m. to 6 p.m. and less from 6 p.m. to midnight.


\subsection{study on measuring \hatsuwaryo}



The rise of online conferencing, facilitated by individual microphones for each participant, has simplified speaker separation in scenarios where each person uses a separate computer. Services supporting online interactions and sales, exemplified by platforms like Gong\footnote{{https://www.gong.io}} and JamRoll\footnote{https://jamroll.poetics-ai.com/}, have experienced significant growth. While online settings allow for straightforward speaker separation, the challenge arises in face-to-face meetings or other environments where multiple voices are intermingled.

Traditionally, technologies utilizing array microphones to estimate voice arrival directions and separate speakers have been explored for such mixed environments. Hylable Inc. has recently commercialized Hylable Discussion\footnote{https://www.hylable.com/products/}, a speech measurement system leveraging array microphones to visualize and quantify speech amounts and rotations for each participant. This system streamlines measurement by allowing the user to set the initial speaker directions.


Additionally, advancements in speaker separation methods have emerged with the introduction of neural networks. An example is Pyannote®, which operates without the need for specialized recording devices like array microphones. Released as an open-source library, Pyannote®\footnote{https://github.com/pyannote/pyannote-audio}\cite{bredin2020pyannote} facilitates easier measurement of \hatsuwaryo{} by providing an alternative to traditional microphone-based methods.


\subsection{study on respiration sensing based on \hukubudousakeisoku}


In the study by Park et al.\cite{respiration_belt}, a breathing sensing experiment was conducted using a pressure sensor mounted on a waist belt. This approach aimed to alleviate the discomfort associated with most breathing sensing methods, which typically involve sensors attached directly to the skin. The experiment involved a seated subject, and notable differences in measured values were reported between normal breathing and breath-holding states. 


On a different front, Kaji et al.\cite{kaji2018wearable} introduced a non-invasive respiration sensing device utilizing ultrasonic transducers. These transducers, commonly employed for organ and fetal monitoring, utilize ultrasonic echoes to detect respiration with greater accuracy than conventional wearable respiration monitoring systems. Despite the sophistication and enhanced accuracy of the device, it is important to mention that it is not commercially available, and details regarding its cost remain undisclosed.

\section{Proposed System}

Next, we present our proposed system designed for measuring \hatsuwazikan{} through \hukubudousakeisokudevice{} equipped with \kuukiatusensor.

\subsection{System overview}

The \hukubudousakeisokudevice, developed in this study, captures \kuukiatsu{} data by incorporating a buffer material filled with \kuukiatusensor{} between the subject's pants and abdomen. The collected data undergoes segmentation, features are extracted from time-separated data, and a machine learning model is employed to measure \hatsuwazikan{} by identifying the presence or absence of speech.

\subsection{Abdominal motion measurement device}

\begin{figure}[t]
 \centering
 \includegraphics[width=0.7\columnwidth]{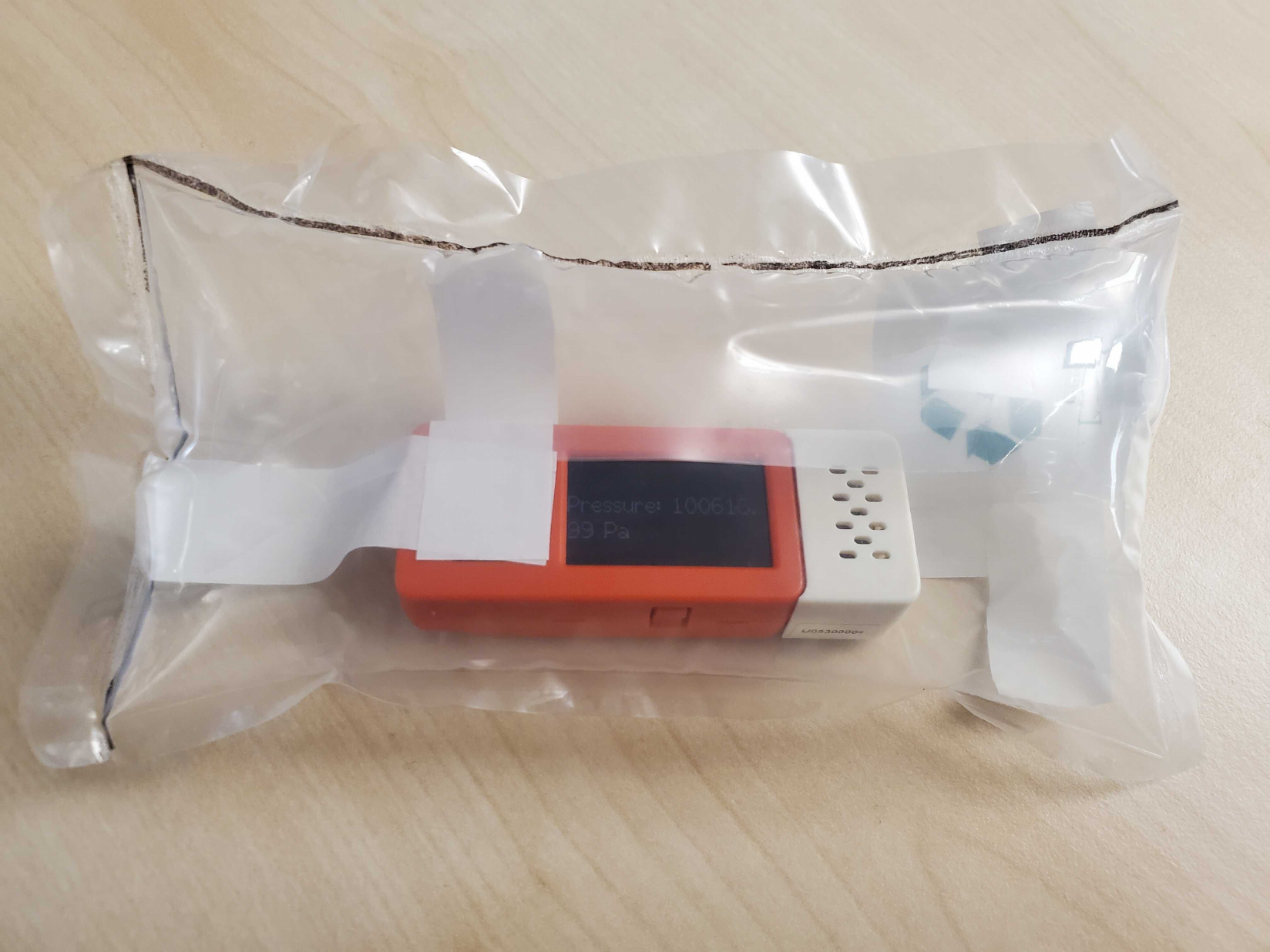}
 \caption{Abdominal motion measurement device}
 \label{fig:device}
\end{figure}
\begin{figure}[t]
 \centering
 \includegraphics[width=0.7\columnwidth]{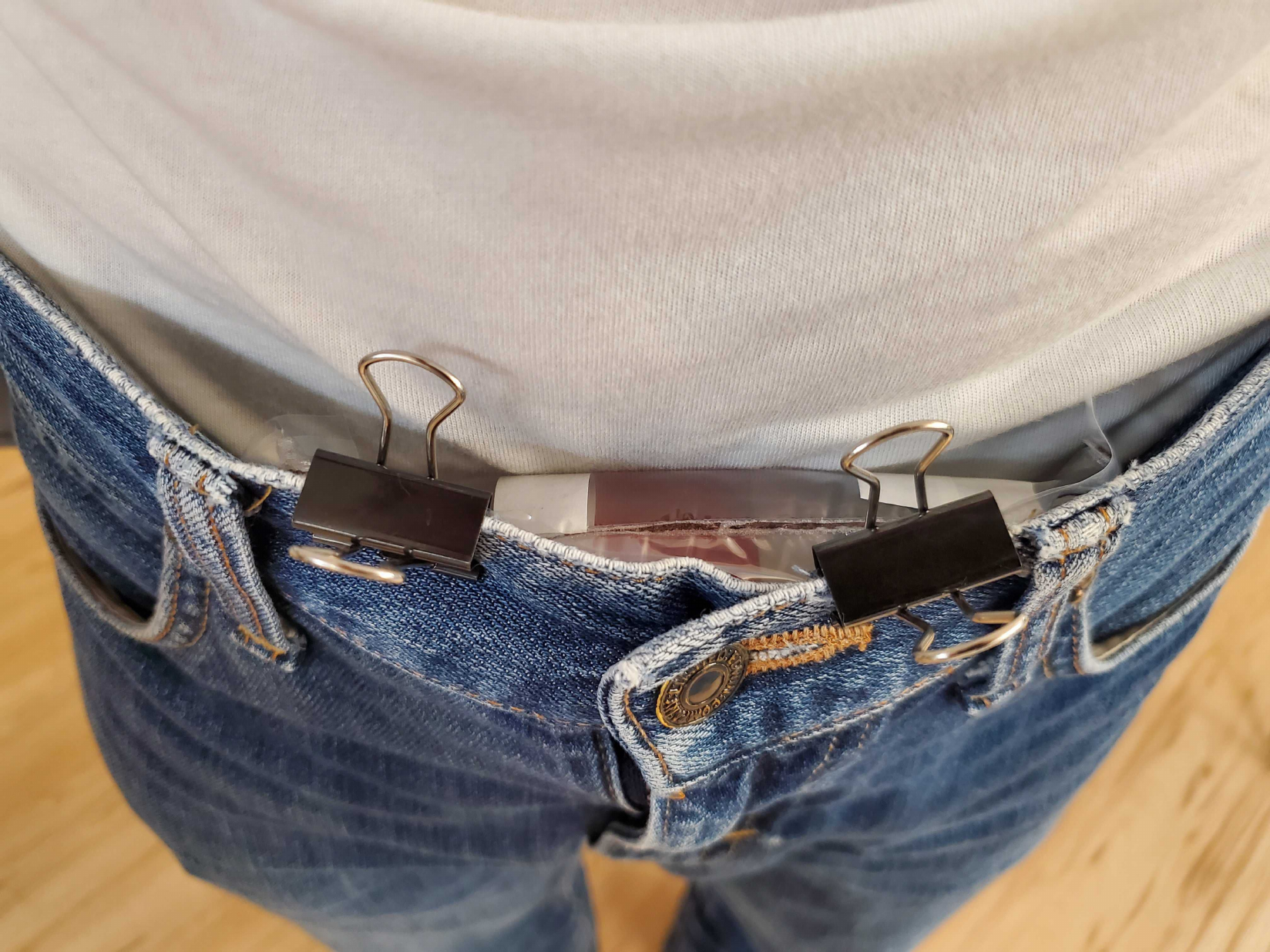}
 \caption{Attachment of abdominal movement measurement device}
 \label{fig:wearing_device}
\end{figure}

The structure of the abdominal motion measurement device utilized in this system is depicted in Fig.\ref{fig:device}. A barometric sensor is enclosed within a buffer material equipped with a check valve. The device measures respiration-induced abdominal motion by monitoring changes in the internal pressure of the buffer material.
The barometric sensor employed is an M5stickC ENV hat from Switch Science, Inc. Due to limitations in the M5stickC ENV hat's ability to measure at higher frequencies, along with inherent time lag in data reception, the sampling frequency is set to approximately 8 Hz.
As illustrated in Fig.\ref{fig:wearing_device}, the device is worn with its center positioned just below the navel. To ensure proper adherence to the abdomen, a belt is used for tightening if necessary.

\subsection{Data Processing}

\begin{figure}[t]
 \centering
 \includegraphics[width=\columnwidth]{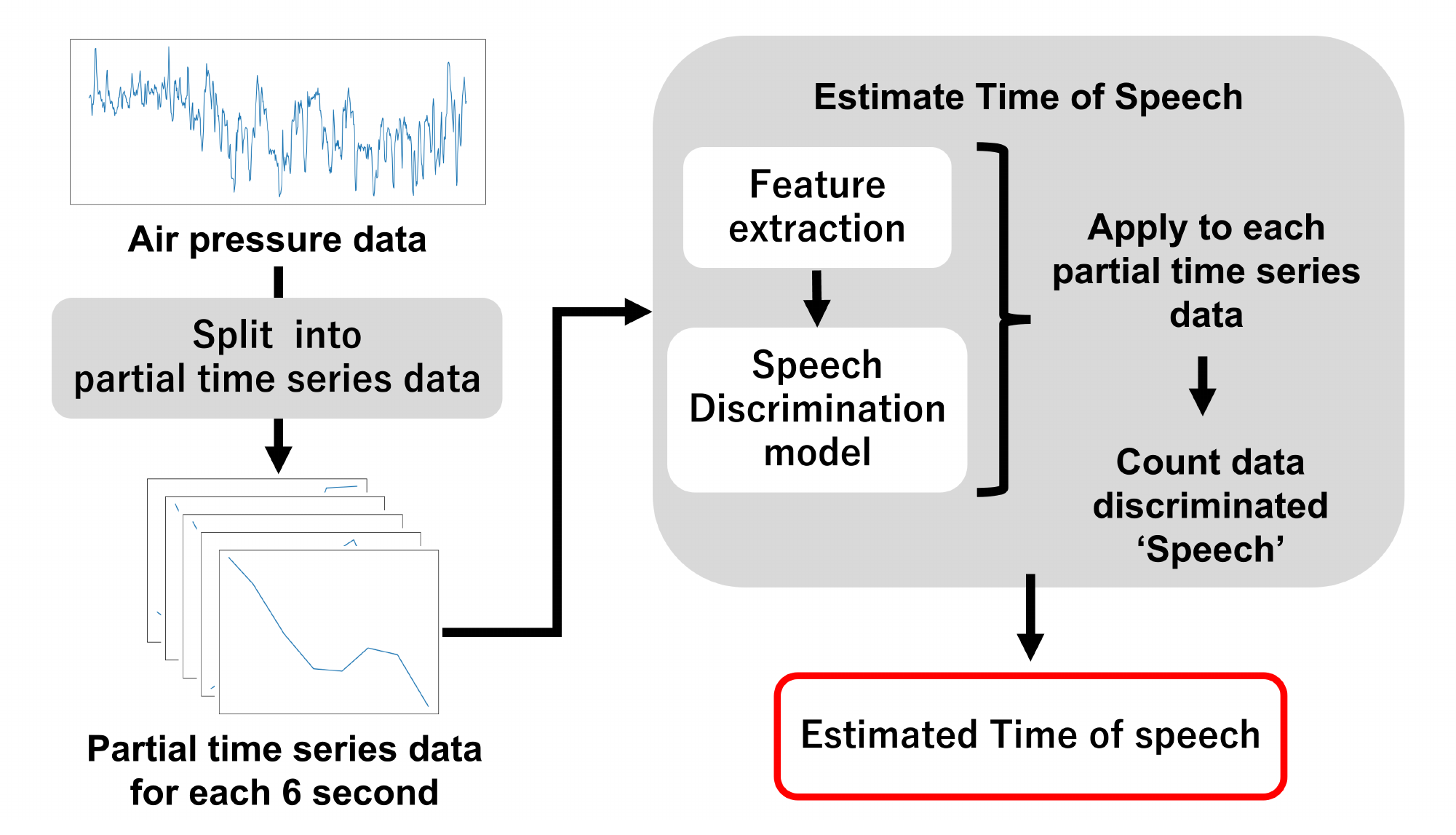}
 \caption{Data processing flow}
 \label{fig:process_data}
\end{figure}

The processing details of the acquired data are elucidated in Fig.\ref{fig:process_data}, with a stepwise explanation provided below.

\subsubsection{Preprocess}

The initially obtained air pressure data underwent a smoothing procedure using rsp\_clean in neuroit2\footnote{https://github.com/neuropsychology/NeuroKit}, a Python Toolbox for Neurophysiological Signal Processing. This step aimed to refine and enhance the quality of \kuukiatsu{} data.

Subsequently, the smoothed \kuukiatsu{} data was subjected to segmentation, resulting in partial time series data. This segmentation involved creating subsets of data with a window width of 6 seconds and a 25 \% overlap, effectively dividing the continuous data into smaller, more manageable segments. This approach facilitates a more granular analysis of the data, enabling precise feature extraction and contributing to the accuracy of subsequent measurements, particularly in the context of \hatsuwazikan{} analysis.

\subsubsection{Feature extraction}

In the feature extraction stage, the following features are derived from \bubunzikeiretudata.

\begin{itemize}
    \item Maximum, Minimum, Mean, Standard deviation, Skewness, Kurtosis in \bubunzikeiretudata
    \item Maximum, Minimum, Mean, Standard deviation, Skewness, Kurtosis, Peak frequency, Average frequency of frequency transformed \bubunzikeiretudata{} by Fast Fourier Transform (FFT)
\end{itemize}


In the case of peak frequency, an attempt was made to adhere to the range of 0.13 Hz to 0.66 Hz, as referenced in the work of Ahmed et al.\cite{freq2023}. If a frequency within this range was not identifiable, the peak frequency was set to zero. These extracted features play a crucial role in characterizing and quantifying the abdominal motion data, contributing to the subsequent steps in the analysis, particularly in the context of speech time measurement.

\subsubsection{Measurement \hatsuwazikan}


In this step, the extracted features are input into a speech discrimination model, which identifies the presence or absence of speech on a per-second basis. The cumulative presence or absence of speech is then aggregated for each 6-second interval, allowing for the measurement of speech time within the recorded timeframe.

Due to the 6-second window size of the data input into the speech discrimination model, the system labels a subject as speaking only if they speak continuously for the entire 6 seconds. However, acknowledging the hypothesis that a discernible difference exists between brief speech instances (lasting less than 6 seconds) and complete silence, the system considers data as speech if spoken for more than 3 seconds out of the 6-second window. Additionally, when the speech discrimination model detects the subject speaking in response to the input data, the duration of speech is treated as 4.5 seconds. This approach is implemented to capture the essence of speech duration within the context of the given model and data window size.

\section{Creation of \hatsuwasikibetumodel}


This section provides a comprehensive account of the creation of \hatsuwasikibetumodel{} within the proposed system.

\subsection{Collection of training data}

In this study, a total of 10 subjects were equipped with the previously described \hukubudousakeisokudevice{} and measured under two distinct conditions: ``\hatsuwanashi'' and ``\hatsuwaari''. The data collected under these conditions were utilized as the training dataset for the subsequent development of \hatsuwasikibetumodel.



Utilizing the M5StickC Plus, which is capable of measuring both air pressure and 3-axis acceleration, the sensor was secured within a buffering material to capture data for both parameters. Subsequently, separate machine learning models were constructed for each dataset, and their accuracies were compared. The sampling frequency for the 3-axis acceleration data matched that of the air pressure data, at approximately 8 Hz.

The experimental procedures for the ``\hatsuwanashi'' and ``\hatsuwaari'' conditions are detailed below.

\subsubsection{\hatsuwanashi}

\begin{figure}[t]
 \centering
 \includegraphics[width=0.7\columnwidth]{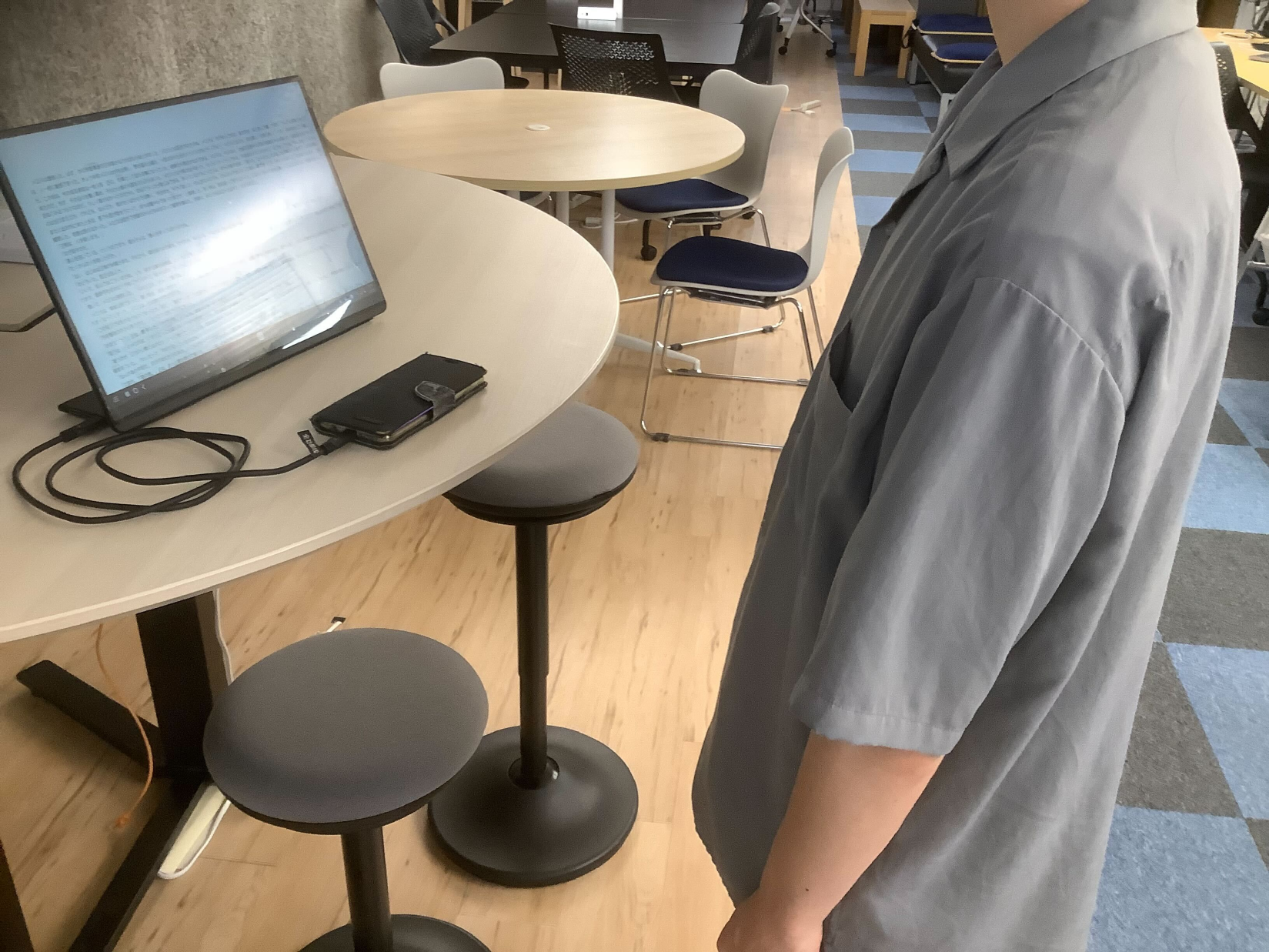}
 \caption{Measurement environment}
 \label{fig:measure_style}
\end{figure}

Subjects were asked to stand naturally in front of the desk, as shown in Fig.\ref{fig:measure_style}. For accurate measurement, subjects were instructed to minimize movement. A monitor was placed in front of their eyes, and a video was played. The subjects watched the video for 90 seconds while their abdominal motion was measured by the device. During the experiment, speaking was prohibited, and only abdominal movements caused by breathing were recorded. Fig.\ref{hatsuwanashi} shows an example of the measurement results.

\subsubsection{\hatsuwaari}

\begin{figure}[t]
 \centering
 \begin{minipage}{0.8\hsize}
  \centering
  \includegraphics[width=\columnwidth]{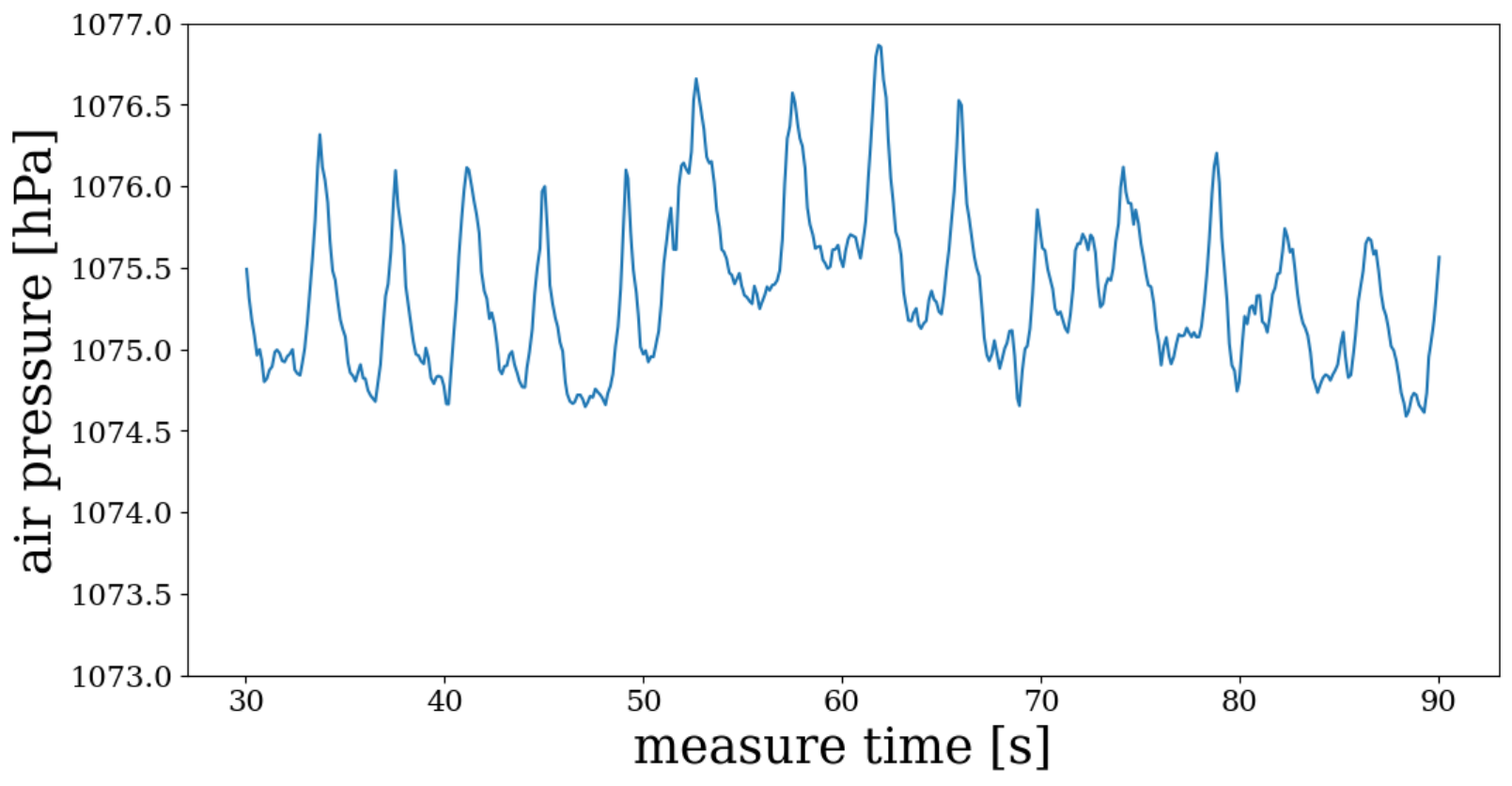} \\
  \subcaption{\hatsuwanashi}
  \label{hatsuwanashi}
 \end{minipage}
 \hfill
 \begin{minipage}{0.8\hsize}
  \centering
  \includegraphics[width=\columnwidth]{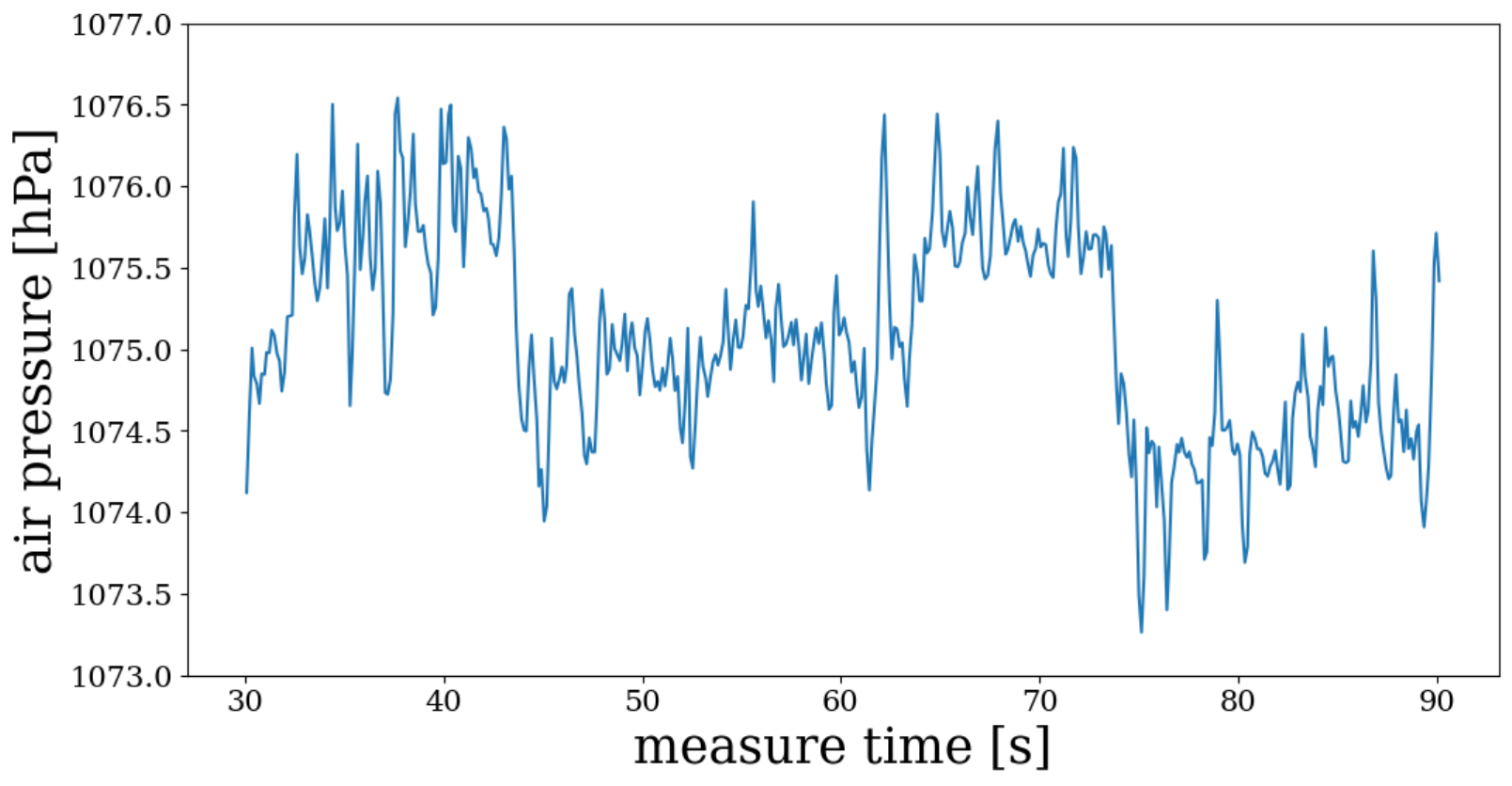} \\
  \subcaption{\hatsuwaari}
  \label{hatsuwaari}
 \end{minipage}
 \caption{Air pressure data}
 \label{fig:fig_pressure}
\end{figure}

Similar to the ``No Speech'' condition, subjects stood naturally in front of the desk. Then, they read aloud the text displayed on the monitor in front of them while their abdominal motion was measured by the device. Subjects were instructed to read aloud at their usual speaking speed and volume as much as possible. An example of the measurement results is shown in Fig. \ref{hatsuwaari}.

\subsection{Data processing}


The collected \kuukiatsu{} data and \sanzikukasokudo{} data spanned a duration of 90 seconds each. However, the initial 30 seconds of data were discarded to allow time for each subject's breathing to stabilize. Subsequently, as outlined in Section 3, features were extracted from the partial time series data, and accurate labels (``\hatsuwanashi'' and ``\hatsuwaari'') were assigned. For \sanzikukasokudo{} data, features were individually extracted for each axis \kasokudo{} and then combined.

This processing resulted in a feature matrix of 891 rows × 14 columns for the air pressure data and a feature matrix of 891 rows × 42 columns for the \sanzikukasokudo{} data. The feature matrices for \kuukiatsu{} and \sanzikukasokudo{} were then concatenated, producing a consolidated feature matrix with dimensions 891 rows × 56 columns. This combined feature matrix serves as the foundation for training and evaluating \hatsuwasikibetumodel.

\subsection{Creation and evaluation of \hatsuwasikibetumodel}


Speech discrimination model were created using three distinct feature matrices: one derived from \kuukiatsu{} data, another from \sanzikukasokudo{} data, and the third by combining air pressure and \sanzikukasokudo{}. Given the absence of prior studies on machine learning models for speech discrimination, three commonly employed supervised learning methods were applied: Support Vector Machine (SVM), Random Forest (RF), and k-Nearest Neighbor (kNN). A search was conducted to identify the model demonstrating the best performance.

Cross-validation was employed with 90 \% of the training data and 10 \% of the validation data. This process was iterated 10 times, and the average accuracy was calculated. The results, as shown in Table \ref{accuracy}, indicated that the highest accuracy for the three training models was achieved when using the Random Forest (RF) method: 89.8 \% for \kuukiatsu{} data, 90.5\% for \sanzikukasokudo{} data, and 94.6 \% for the combined \kuukiatsu{} and \sanzikukasokudo{} data.

It is noteworthy that the RF model outperformed the other methods across all feature matrices. The accuracy of the training model using \kuukiatsu{} data was slightly lower than that using \sanzikukasokudo{} data. Potential reasons for this discrepancy are elaborated below.

\begin{table}[htbp]
\caption{Evaluation of \hatsuwasikibetumodel}
\begin{center}
\renewcommand{\arraystretch}{1.2}{
\begin{tabular}{c|c|c|c}
\hline
\textbf{\textit{model}}& \textbf{\textit{\kuukiatsu}}& \textbf{\textit{\sanzikukasokudo}}& \textbf{\textit{both data}} \\
\hline
\textbf{\textit{RF}}& \textbf{89.8 \%}& \textbf{90.5 \%}& \textbf{94.6 \%} \\
\hline
\textbf{\textit{SVM}}& {88.0 \%}& {78.3 \%}& {91.0 \%} \\
\hline
\textbf{\textit{kNN}}& {87.6 \%}& {80.3 \%}& {90.5 \%}\\
\hline
\end{tabular}
}
\label{accuracy}
\end{center}
\end{table}

\subsubsection{Posture Fixation}

In the experiments where the training data were collected, subjects were measured in a standing position, and their movements were minimal. As mentioned in Section 2, one challenge with acceleration-based breathing measurements is their susceptibility to motion artifacts. However, the methodology employed in this study was not influenced by the subject's motion, contributing to the robustness of the measurements.

\subsubsection{Sampling frequency}


Both \kuukiatsu{} and \sanzikukasokudo{} data were sampled at approximately 8 Hz in this study. The window width was set to 6 seconds to facilitate adequate frequency analysis. However, it is essential to note that this aspect is sensor-dependent because, for accurate speech time measurement, the window width needs to be as small as possible.


Based on these considerations and the achieved results, the Random Forest (RF) algorithm was selected for all three \hatsuwasikibetumodel{}. The model based on \kuukiatsu{} data is referred to as Model 1, the model based on \sanzikukasokudo{} data is Model 2, and the model based on both data is Model 3. The adoption of the RF algorithm for all models underscores its effectiveness in discerning speech patterns from the provided data.

\section{Experiments to validate \hatsuwasikibetumodel}


Next, we will verify and evaluate the applicability of the three speech discrimination models created in Section 4 to estimating speech time using data collected from real-life conversations.

\subsection{Data collection}

Three subjects (ID 1-3) were instructed to engage in speech during a weekly meeting conducted in our laboratory. Similar to the procedure for collecting training data, subjects assumed a standing position, with the abdominal motion measurement device attached to their pants and abdomen. Over a period of 20 minutes, air pressure and acceleration data were recorded.


Furthermore, instances involving aids and brief responses (e.g., ``yes'' or ``I see'') were excluded from the calculation of speech duration. Additionally, any stuttering occurring during speech was considered part of the speech duration unless the stuttering persisted for more than 3 seconds.


\subsection{Estimated time of speech}


For the collected air pressure and 3-axis acceleration data, we generated partial time series data segments with a window width of 6 seconds. Feature extraction was then performed as described in Section 3. The resulting speech features were fed into the speech discrimination model created in Section 4, enabling speech identification every 6 seconds. We then counted the data segments identified as ``Speech'' to estimate the total speech duration in seconds.

\begin{figure}[t]
 \centering
 \includegraphics[width=0.7\columnwidth]{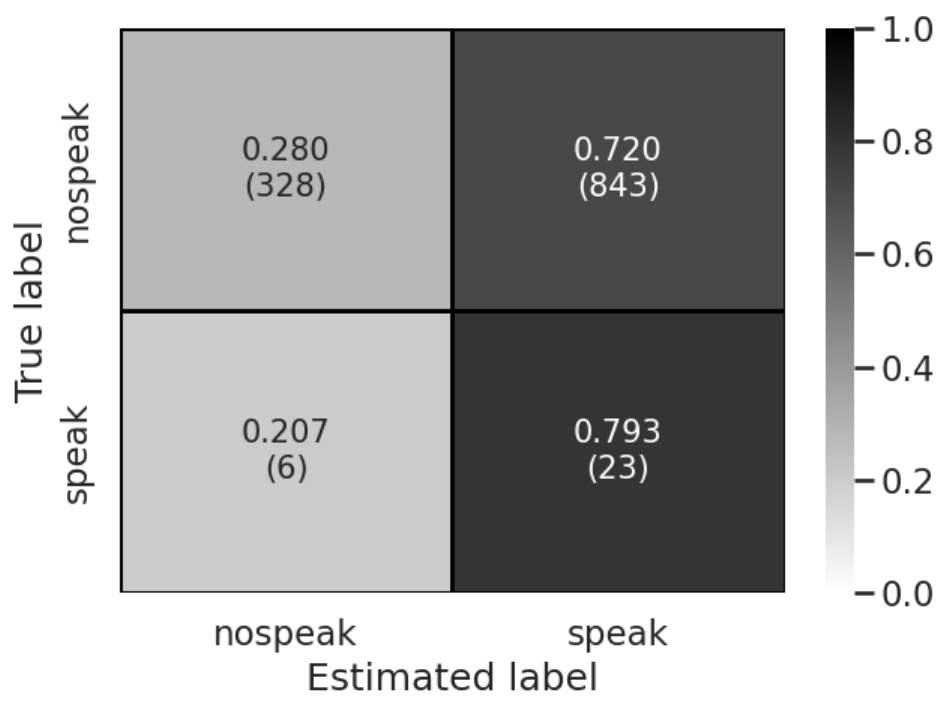}
 \caption{confusion matrix for Model1 (ID1)}
 \label{fig:fig_estimated_1}
\end{figure}

Table \ref{speak_time} illustrates the actual speech time for each subject alongside the estimated speech duration by the speech discrimination model. In all cases, the actual speech time considerably exceeded the estimated speech duration. Furthermore, Fig.\ref{fig:fig_estimated_1} presents a confusion matrix depicting the predicted and correct labels, indicating a predominant prediction class of ``Speech''. Video recordings reveal that during most of the measurement time, the three subjects were engaged in computer operations. Additionally, the subjects were instructed to maintain comfortable postures throughout the 20-minute standing measurement.

The subjects' posture during the measurement exhibited a slight stoop, differing significantly from the measurement environment in Section 4 where training data were collected. The proposed method in this study may introduce variations in the device's wearing position among subjects. Despite efforts to clearly define the device's wearing position, differences in body size and shifts in the measurement device's position during the experiment in Section 4 contributed to variations in measurement data among subjects, affecting the results.

\begin{table*}[htbp]
\caption{Performance of speech discrimination model using speech data during meetings}
\begin{center}
\renewcommand{\arraystretch}{1.2}{
\begin{tabular}{c|c|c|c|c}
\hline
\cline{2-4} 
\textbf{\textit{ID}}& \textbf{\textit{actual time [s]}}& \textbf{\textit{estimated time (Model 1) [s]}}& \textbf{\textit{estimated time (Model 2) [s]}}& \textbf{\textit{estimated time (Model 3) [s]}} \\
\hline
\textbf{\textit{1}}& {29}& {459}& {445.5} & {468}\\
\hline
\textbf{\textit{2}}& {197}& {603}& {747} & {621}\\
\hline
\textbf{\textit{3}}& {221}& {396}& {652.5} & {490.5}\\
\hline
\end{tabular}
}
\label{speak_time}
\end{center}
\end{table*}

\section{Conclusion}
Finally, this section describes the results of this study and future prospects.

\subsection{Results of this study}


This study aimed to develop a cost-effective method for measuring daily speech production while addressing privacy concerns. We focused on analyzing respiratory patterns during speech and non-speech intervals. Abdominal movements were quantified using a barometric sensor. A speech discrimination model was then developed based on this data to estimate speech duration from actual speech recordings.We proposed and implemented a method employing a barometric sensor to measure abdominal movements. The abdominal motion measurement device, utilizing the barometric sensor, captured abdominal motion by monitoring changes in internal air pressure.


The accuracy of the speech discrimination model relying on air pressure data reached 89.8\% (detailed in Section 4). Similarly, the speech discrimination model based on 3-axis acceleration data achieved an accuracy of 90.5\%, demonstrating the feasibility of measuring speech duration without a microphone.However, during the evaluation of speech duration in a conversational setting (Section 5), particularly in meetings, the estimated speech time significantly exceeded the actual time. This discrepancy was attributed to substantial changes in posture throughout the measurement.

\subsection{Future work}


This study underscores the significant impact of posture on speech measurement without a microphone. Future studies should focus on developing a measurement method that remains unaffected by posture, regardless of whether individuals are walking, sitting, or standing. Alternatively, a device or system that adjusts the analysis method based on posture variations may be necessary. Traditionally, acceleration sensors have been used to detect posture and have been integrated into devices from previous studies. However, this study chose a barometric sensor for the abdominal motion measurement device because acceleration can be easily influenced by motion.

\section*{Acknowledgment}
This study was supported in part by Grant-in-Aid for Scientific study (JP19H05665) and by the study Center Support Program for Society 5.0 (JPMXP0518071489).

\bibliographystyle{IEEEtran} 

\end{document}